\documentclass[a4paper]{eptcs}
\usepackage[english]{babel}
\usepackage{hyperref}
\usepackage{appendix}		
\usepackage{listings}
\usepackage[usenames,dvipsnames]{color}
\usepackage{graphicx, subfig}
	\DeclareGraphicsExtensions{.pdf,.png,.jpg,.jpeg,.mps,.emf}
\usepackage{epstopdf}
\usepackage{multirow}
\usepackage{xspace}	
\usepackage{mdwlist}
\usepackage{prooftree}
\usepackage{amsmath, amsthm, amssymb,xcolor,wasysym, altbeginend}
\usepackage{comment}		
\usepackage{mathptmx,latexsym }
\usepackage{stmaryrd}
\usepackage{eurosym}
\usepackage{tabularx}
\usepackage{wrapfig}
\usepackage{breakurl}
\usepackage{enumitem}
\setlist{nolistsep}

\hypersetup{
  pdftitle={Multiparty Session Actors},
  pdfauthor={Rumyana Neykova and Nobuko Yoshida},
  pdfsubject={Multiparty Session Actors},
  pdfkeywords={Python, session types, Scribble, actors},
  colorlinks=true,	
  linkcolor	= black, 
  citecolor = black,
  urlcolor = black
}

\begin{document}
%

%

%
%
%
%
%
%

\makeatletter

%
\let\if@envcntreset\iffalse
\let\if@envcntsame\iffalse
\let\if@envcntsect\iftrue

%
\def\@thmcountersep{}
\def\@thmcounterend{.}

\def\spnewtheorem{\@ifstar{\@sthm}{\@Sthm}}


\def\@spnthm#1#2{%
  \@ifnextchar[{\@spxnthm{#1}{#2}}{\@spynthm{#1}{#2}}}
\def\@Sthm#1{\@ifnextchar[{\@spothm{#1}}{\@spnthm{#1}}}

\def\@spxnthm#1#2[#3]#4#5{\expandafter\@ifdefinable\csname #1\endcsname
   {\@definecounter{#1}\@addtoreset{#1}{#3}%
   \expandafter\xdef\csname the#1\endcsname{\expandafter\noexpand
     \csname the#3\endcsname \noexpand\@thmcountersep \@thmcounter{#1}}%
   \expandafter\xdef\csname #1name\endcsname{#2}%
   \global\@namedef{#1}{\@spthm{#1}{\csname #1name\endcsname}{#4}{#5}}%
                              \global\@namedef{end#1}{\@endtheorem}}}

\def\@spynthm#1#2#3#4{\expandafter\@ifdefinable\csname #1\endcsname
   {\@definecounter{#1}%
   \expandafter\xdef\csname the#1\endcsname{\@thmcounter{#1}}%
   \expandafter\xdef\csname #1name\endcsname{#2}%
   \global\@namedef{#1}{\@spthm{#1}{\csname #1name\endcsname}{#3}{#4}}%
                               \global\@namedef{end#1}{\@endtheorem}}}

\def\@spothm#1[#2]#3#4#5{%
  \@ifundefined{c@#2}{\@latexerr{No theorem environment `#2' defined}\@eha}%
  {\expandafter\@ifdefinable\csname #1\endcsname
  {\global\@namedef{the#1}{\@nameuse{the#2}}%
  \expandafter\xdef\csname #1name\endcsname{#3}%
  \global\@namedef{#1}{\@spthm{#2}{\csname #1name\endcsname}{#4}{#5}}%
  \global\@namedef{end#1}{\@endtheorem}}}}

\def\@spthm#1#2#3#4{\topsep 7\p@ \@plus2\p@ \@minus4\p@
\refstepcounter{#1}%
\@ifnextchar[{\@spythm{#1}{#2}{#3}{#4}}{\@spxthm{#1}{#2}{#3}{#4}}}

\def\@spxthm#1#2#3#4{\@spbegintheorem{#2}{\csname the#1\endcsname}{#3}{#4}%
                    \ignorespaces}

\def\@spythm#1#2#3#4[#5]{\@spopargbegintheorem{#2}{\csname
       the#1\endcsname}{#5}{#3}{#4}\ignorespaces}

\def\@spbegintheorem#1#2#3#4{\trivlist
                 \item[\hskip\labelsep{#3#1\ #2\@thmcounterend}]#4}

\def\@spopargbegintheorem#1#2#3#4#5{\trivlist
      \item[\hskip\labelsep{#4#1\ #2}]{#4(#3)\@thmcounterend\ }#5}


\def\@sthm#1#2{\@Ynthm{#1}{#2}}

\def\@Ynthm#1#2#3#4{\expandafter\@ifdefinable\csname #1\endcsname
   {\global\@namedef{#1}{\@Thm{\csname #1name\endcsname}{#3}{#4}}%
    \expandafter\xdef\csname #1name\endcsname{#2}%
    \global\@namedef{end#1}{\@endtheorem}}}

\def\@Thm#1#2#3{\topsep 7\p@ \@plus2\p@ \@minus4\p@
\@ifnextchar[{\@Ythm{#1}{#2}{#3}}{\@Xthm{#1}{#2}{#3}}}

\def\@Xthm#1#2#3{\@Begintheorem{#1}{#2}{#3}\ignorespaces}

\def\@Ythm#1#2#3[#4]{\@Opargbegintheorem{#1}
       {#4}{#2}{#3}\ignorespaces}

\def\@Begintheorem#1#2#3{#3\trivlist
                           \item[\hskip\labelsep{#2#1\@thmcounterend}]}

\def\@Opargbegintheorem#1#2#3#4{#4\trivlist
      \item[\hskip\labelsep{#3#1}]{#3(#2)\@thmcounterend\ }}

\if@envcntsect
   \def\@thmcountersep{.}
   \spnewtheorem{theorem}{Theorem}[section]{\bfseries}{\itshape}
\else
   \spnewtheorem{theorem}{Theorem}{\bfseries}{\itshape}
   \if@envcntreset
      \@addtoreset{theorem}{section}
   \else
      \@addtoreset{theorem}{chapter}
   \fi
\fi

\spnewtheorem*{claim}{Claim}{\itshape}{\rmfamily}
\if@envcntsame 
   \def\spn@wtheorem#1#2#3#4{\@spothm{#1}[theorem]{#2}{#3}{#4}}
\else 
   \if@envcntsect 
      \def\spn@wtheorem#1#2#3#4{\@spxnthm{#1}{#2}[section]{#3}{#4}}
   \else 
      \if@envcntreset
         \def\spn@wtheorem#1#2#3#4{\@spynthm{#1}{#2}{#3}{#4}
                                   \@addtoreset{#1}{section}}
      \else
         \def\spn@wtheorem#1#2#3#4{\@spynthm{#1}{#2}{#3}{#4}
                                   \@addtoreset{#1}{chapter}}%
      \fi
   \fi
\fi
\spn@wtheorem{case}{Case}{\itshape}{\rmfamily}
\spn@wtheorem{conjecture}{Conjecture}{\itshape}{\rmfamily}
\spn@wtheorem{corollary}{Corollary}{\bfseries}{\itshape}
\spn@wtheorem{definition}{Definition}{\bfseries}{\itshape}
\spn@wtheorem{example}{Example}{\itshape}{\rmfamily}
\spn@wtheorem{exercise}{Exercise}{\itshape}{\rmfamily}
\spn@wtheorem{lemma}{Lemma}{\bfseries}{\itshape}
\spn@wtheorem{problem}{Problem}{\itshape}{\rmfamily}
\spn@wtheorem{property}{Property}{\itshape}{\rmfamily}
\spn@wtheorem{proposition}{Proposition}{\bfseries}{\itshape}
\spn@wtheorem{question}{Question}{\itshape}{\rmfamily}
\spn@wtheorem{solution}{Solution}{\itshape}{\rmfamily}
\spn@wtheorem{remark}{Remark}{\itshape}{\rmfamily}

\def\@takefromreset#1#2{%
    \def\@tempa{#1}%
    \let\@tempd\@elt
    \def\@elt##1{%
        \def\@tempb{##1}%
        \ifx\@tempa\@tempb\else
            \@addtoreset{##1}{#2}%
        \fi}%
    \expandafter\expandafter\let\expandafter\@tempc\csname cl@#2\endcsname
    \expandafter\def\csname cl@#2\endcsname{}%
    \@tempc
    \let\@elt\@tempd}

\def\theopargself{\def\@spopargbegintheorem##1##2##3##4##5{\trivlist
      \item[\hskip\labelsep{##4##1\ ##2}]{##4##3\@thmcounterend\ }##5}
                  \def\@Opargbegintheorem##1##2##3##4{##4\trivlist
      \item[\hskip\labelsep{##3##1}]{##3##2\@thmcounterend\ }}
      }

\makeatother

\newcommand{\MYPARAGRAPH}[1]{\paragraph{\textnormal{\textbf{#1}}}}

\definecolor{dkblue}{rgb}{0,0.1,0.5}
\definecolor{dkgreen}{rgb}{0,0.4,0}
\definecolor{dkred}{rgb}{0.4,0,0}

\newcommand{\CODESIZE}{\small}
\newcommand{\CODESTYLE}{\ttfamily}

\lstset%
{%
	captionpos=b,
	columns=flexible, 
	commentstyle=\CODESTYLE\footnotesize\color{purple},
	escapeinside={*@}{@*},
	float=hbp,
	frame=none,
	language=Java,
	mathescape=true,
	numbers=none, 
	numberstyle=\tiny,
	showspaces=false,
	showstringspaces=false,
	showtabs=false,
	stringstyle=\color{teal},
	tabsize=2
}

\newcommand{\CODE}[1]{\texttt{\CODESIZE#1}} 
\newcommand{\GREYCODE}[1]{\CODE{
	\color{black}{#1}}} 
\newcommand{\SJCODE}[1]
{%
	\lstinline[style=SJ]+#1+%
}

\lstnewenvironment{CODELISTING}
{%
	\onehalfspacing
	\lstset%
	{%
		basicstyle=\CODESTYLE\footnotesize,
		keywordstyle=\CODESTYLE\footnotesize,
	}
}%
{%
	\doublespacing 
}

\lstdefinestyle{SJ}%
{%
	basicstyle   = \CODESTYLE\footnotesize,
	keywordstyle=[1]{
	\!\!\!\color{dkblue}
	\CODESTYLE\footnotesize},
	keywordstyle=[2]{
	\!\!\!\color{dkgreen}
	\CODESTYLE\footnotesize},
        moredelim=*[s][\footnotesize\color{dkgreen}]{<}{>},
        morekeywords =
	[1]{%
		protocol, role, choice, at, or, from, to, rec, parallel,
                and, interrupt, by, finish, continue, global, local, self, interruptible, with
	},
	morekeywords =
	[2]{%
                int,Data,
	},
       literate={>=}{$\geq\ $}{2}{<=}{$\leq\ $}{2}
}

\lstnewenvironment{SJLISTING}%
{
	\lstset{style=SJ}
}
{
}

%
\newcommand{\REF}[1]{\S\,\ref{#1}}

%
\newcommand{\OPASSIGN}{\, \CODE{:=} \,}
\newcommand{\OPEQ}{\ensuremath{=}}
\newcommand{\OPDEC}{\CODE{--}}
\newcommand{\OPINC}{\CODE{++}}
\newcommand{\OPSEQ}{\CODE{;}}
\newcommand{\OPNOT}{\ensuremath{\neg}}

\newcommand{\OLINE}[1]{\ensuremath{\overline{#1}}}  
\newcommand{\SET}[1]{\ensuremath{\{ #1 \}}}         
\newcommand{\FUN}[2]{\ensuremath{\mathsf{#1}(#2)}}  

%
\newcommand{\KWORD}[1]{\ensuremath{\mathsf{#1}}}
\newcommand{\DTYPE}[1]{\ensuremath{\mathtt{#1}}}  
\newcommand{\DVAL}[1]{\ensuremath{\mathtt{#1}}}   
\newcommand{\PPAR}{\ensuremath{\, | \,}}
\newcommand{\PPARGROUP}[1]{\ensuremath{(#1)}}
\newcommand{\PIF}{\ensuremath{\KWORD{if}}}
\newcommand{\PTHEN}{\ensuremath{\KWORD{then}}}
\newcommand{\PELSE}{\ensuremath{\KWORD{else}}}
\newcommand{\PNIL}{\ensuremath{\mathbf{0}}}
\newcommand{\EMPTY}{\ensuremath{\epsilon}}

%
\newcommand{\LAB}[1]
	{\ensuremath{#1}}            
\newcommand{\LABVAL}[1]
	{\ensuremath{\mathtt{#1}}}   
\newcommand{\ROLE}[1]
	{\participant{#1}}           
\newcommand{\PARTY}[1]
	{\ensuremath{\mathsf{#1}}}   
\newcommand{\MSGlxS}[3]
	{\ensuremath{\MSGlx{#1}{#2\!:\!#3}}}  
\newcommand{\MSGlx}[2]
	{\ensuremath{#1 (#2)}}              
\newcommand{\MSGlS}[2]
	{\MSGlx{#1}{#2}}                    
\newcommand{\BRANCH}[1]
	{\ensuremath{\SET{#1}}}         
\newcommand{\MUREC}[1]
	{\ensuremath{\mu \RECVAR{#1}}}  
\newcommand{\RECVAR}[1]
	{\ensuremath{\keyword{#1}}}     

%
\newcommand{\STATEVAR}[1]
	{\ensuremath{\mathtt{#1}}}                
\newcommand{\ROLEVAR}[2]
	{\ensuremath{\ROLE{#1} . \STATEVAR{#2}}}  
\newcommand{\STATEVARDECL}[2]
	{\ensuremath{\STATEVAR{#1} : \DTYPE{#2}}}      
\newcommand{\PARTYSTATEDECL}[2]
	{\ensuremath{\ROLE{#1}} : [#2]}                

%
\newcommand{\GSEP}{\ensuremath{.}}  
\newcommand{\GLOBAL}[1]
	{\ensuremath{\mathcal{#1}}}       
\newcommand{\GLOBALi}[2]
	{\ensuremath{\GLOBAL{#1}_{#2}}}   
\newcommand{\GSEND}[2]
	{\ensuremath{\ROLE{#1} \rightarrow \ROLE{#2} :}}  

\newcommand{\GBRA}[1]
	{\ensuremath{\SET{#1}}}         
\newcommand{\GREC}[1]
	{\ensuremath{\mu \RECVAR{#1}}}  

%
\newcommand{\GLOBALDECL}[1]
	{\ensuremath{((#1))}}             
\newcommand{\LASS}[1]
	{\ensuremath{\langle #1 \rangle}}  
\newcommand{\LEFF}[1]
	{\LASS{#1}}                        
\newcommand{\LASSEFF}[2]
	{\ensuremath{\LASS{#1,\, #2}}}     
\newcommand{\RASS}[1]
	{\ensuremath{\{ #1 \}}}            
\newcommand{\REFF}[1]
	{\RASS{#1}}                        
\newcommand{\RASSEFF}[2]
	{\ensuremath{\RASS{#1,\, #2}}}     
\newcommand{\GRECtexA}[4]
	{\ensuremath{\MUREC{#1} \langle #2 \rangle (#3) \{ #4 \}}}
\newcommand{\GRECVARte}[2]
	{\ensuremath{\RECVAR{#1} \langle #2 \rangle}}

%
\newcommand{\LSEP}{\ensuremath{.}}  
\newcommand{\LOCAL}[1]
	{\ensuremath{\mathcal{#1}}}       
\newcommand{\LOCALi}[2]
	{\ensuremath{\LOCAL{#1}_{#2}}}    
\newcommand{\LSEND}[1]
	{\ensuremath{\ROLE{#1} \,!\,}}    
\newcommand{\LRECV}[1]
	{\ensuremath{\ROLE{#1} \,?}}      

%
\newcommand{\LOCALDECL}[1]
	{\ensuremath{[#1]}}               
\newcommand{\LRECtexA}[4]
	{\GRECtexA{#1}{#2}{#3}{#4}}
\newcommand{\LRECVARte}[2]
	{\GRECVARte{#1}{#2}}

%
\newcommand{\POSEP}{\ensuremath{;}}  
\newcommand{\PESEP}{\ensuremath{;}}  
\newcommand{\PSEP}{\ensuremath{.}}   
\newcommand{\PINIT}[1]
	{\ensuremath{\mathsf{#1}}}         
\newcommand{\PINITi}[2]
	{\ensuremath{\mathsf{#1}_{#2}}}    
\newcommand{\PREQ}[4]
	{\ensuremath{\OLINE{#1} \langle #2 [\ROLE{#3}] : \GLOBAL{#4} \rangle}}
\newcommand{\PACC}[4]
	{\ensuremath{#1 ( #2 [ \ROLE{#3} ] : \GLOBAL{#4} )}}          
\newcommand{\PSEND}[5]
	{\ensuremath{#1 [\ROLE{#2}, \ROLE{#3}] \,!\, \LAB{#4} \langle #5 \rangle}}
\newcommand{\PRECV}[3]
	{\ensuremath{#1 [\ROLE{#2}, \ROLE{#3}] \,?\,}}                
\newcommand{\PRECX}[1]
	{\ensuremath{\mu #1}}                            
\newcommand{\PRECXx}[2]
	{\ensuremath{\PRECX{#1} (#2)}}                   
\newcommand{\PRECe}[1]
	{\ensuremath{\langle #1 \rangle}}                
\newcommand{\PRECVARX}[1]
	{\ensuremath{#1}}                                
\newcommand{\PRECVARXe}[2]
	{\ensuremath{\PRECVARX{#1} \langle #2 \rangle}}  

\newcommand{\PRECx}[2]
	{\ensuremath{\MUREC{#1} (#2)}}
\newcommand{\PRECtx}[2]
	{\ensuremath{\MUREC{#1} (#2)}}      
\newcommand{\PRECVARte}[2]
	{\GRECVARte{#1}{#2}}

%
\newcommand{\PNEWKW}{\KWORD{new}}
\newcommand{\PREGKW}{\KWORD{reg}}
\newcommand{\PINKW}{\KWORD{in}}
\newcommand{\PNEWs}[4]
	{\ensuremath{\PNEWKW\, (\AT{#1}{#2}, \ROLE{#3}) \,\PINKW\, #4}}  
\newcommand{\PNEWa}[4]
	{\ensuremath{\PREGKW\, \AT{#1}{#2}[\ROLE{#3}] \,\PINKW\, #4}}
\newcommand{\PNEWp}[4]
	{\ensuremath{\PNEWKW\, \AT{#1}{#2} \,\KWORD{with}\, [#3] \,\PINKW\, #4}}

\newcommand{\PJOIN}[2]
	{\ensuremath{\KWORD{join}\, #1[#2]}}  

%
\newcommand{\PLOCK}
	{\ensuremath{\blacktriangledown \,}}
\newcommand{\PUNLOCK}
	{\ensuremath{\blacktriangle}}
\newcommand{\PGET}[2]
	{\ensuremath{#1 \OPASSIGN get(\STATEVAR{#2})}}  
\newcommand{\PPUT}[2]
	{\ensuremath{put(#1, \STATEVAR{#2})}}           
\newcommand{\PLRECV}[3]
	{#1 [\ROLE{#2}, \ROLE{#3}] \, ? \blacktriangledown}

%
\newcommand{\PSNET}[1]
	{\ensuremath{#1}}     
\newcommand{\PSNETi}[2]
	{\ensuremath{#1_{#2}}}
\newcommand{\PICHAN}[2]
	{\ensuremath{\mathtt{I}(#1 [\ROLE{#2}])}}
\newcommand{\POCHAN}[2]
	{\ensuremath{\mathtt{O}(#1 [\ROLE{#2}])}}
\newcommand{\PNETQUEUE}[2]
	{\ensuremath{\langle #1 ; #2 \rangle}}

\newcommand{\RAYCOMMENT}[1]{~\\ \textbf{RAY:} #1}

\newcommand{\capabilities}{\mathtt{c}}
\newcommand{\GQueue}{h}
\newcommand{\monitorSet}{D}
\newcommand{\inTop}{\text{ in }}
\newcommand{\scribble}{Scribble}
\newcommand{\java}{{{\sc Java}}}
\newcommand{\eval}{\downarrow}
\newcommand{\ocaml}{{{\sc Ocaml}}}
\newcommand{\mN}{\mathsf{N}}

    \newcommand{\com}[2]{\par
      \fcolorbox{red}{yellow}{\parbox{\linewidth}{ 
            \color{gray}
            \begin{description}
            \item[{\color{blue} #2:}]{\sf #1}
            \end{description}}}
    }

\spnewtheorem{DEF}[theorem]{Definition}{\bfseries}{\rmfamily}
\spnewtheorem{REM}[theorem]{Remark}{\bfseries}{\rmfamily}
\spnewtheorem{PRO}[theorem]{Proposition}{\bfseries}{\rmfamily}
\spnewtheorem{CON}[theorem]{Convention}{\bfseries}{\rmfamily}
\spnewtheorem{LEM}[theorem]{Lemma}{\bfseries}{\itshape}
\spnewtheorem{THM}[theorem]{Theorem}{\bfseries}{\itshape}
\spnewtheorem{COR}[theorem]{Corollary}{\bfseries}{\itshape}
\spnewtheorem{EX}[theorem]{Example}{\bfseries}{\rmfamily}

\newcommand{\assleft}{\llbracket}
\newcommand{\assright}{\rrbracket}
\newcommand{\INTp}[1]{\II{#1}}
\newcommand{\OUTp}[1]{\OO{#1}}
\newcommand{\PENDINV}[4]{\{\AT{#1}{#2}[#3] \}_{#4}}

\newcommand{\NI}{\noindent}
\newcommand{\CD}{\!\cdot\!}
\newcommand{\CAL}[1]{\mathcal{#1}}
\newcommand{\OL}[1]{\overline{#1}}
\newcommand{\DIFF}{\backslash}
\newcommand{\Dropp}[2]{#1-#2}
\newcommand{\ifthenelse}{}
\newcommand{\VEC}{\tilde}
\newcommand{\VECw}{\widetilde}

\newcommand{\ENCan}[1]{\langle #1 \rangle}
\newcommand{\ENCda}[1]{\langle\!\langle #1 \rangle\!\rangle}
\newcommand{\ENCdp}[1]{(\!( #1 )\!)}

\newcommand{\ASET}[1]{\{#1\}}
\newcommand{\PAR}{\mathrel{\mid}}
\newcommand{\AT}[2]{#1\! : \! #2}

\newcommand{\OP}[2]{#1\ENCan{#2}}
\newcommand{\OT}[2]{#1(#2)} 
\newcommand{\MSG}[4]{\ENCan{#1, #2, \OP{#3}{#4}}}
\newcommand{\MSGT}[4]{\ENCan{#1, #2, \OT{#3}{#4}}}

\newcommand{\LT}{T} 

\newcommand{\GAbody}{\mathcal{H}}  
\newcommand{\GAssert}{G}  
\newcommand{\GA}{\GAssert}
\newcommand{\LAssert}{T}  
\newcommand{\LA}{\LAssert}
\newcommand{\LAbody}{\mathcal{U}}
\newcommand{\sort}{(\LAssert[\player])}
\newcommand{\MT}{\mathnormal{mv}} 
\newcommand{\AssertEnv}{\Gamma^A} 
\newcommand{\REFINES}{\Supset}
\newcommand{\CATCH}{~\mathtt{catch}~}
\newcommand{\CATCHAT}[2]{\CATCH\mathtt{at}~#1~\mathtt{to}~#2}

\newcommand{\mode}[1]{\keyword{m}(#1)}
\newcommand{\Imode}{\mathtt{I}}
\newcommand{\Omode}{\mathtt{O}}
\newcommand{\IOmode}{\mathtt{IO}}
\newcommand{\imode}{\mathsf{}}
\newcommand{\omode}{\mathsf{}}

\newcommand{\rro}{\ROLE{r}_1}
\newcommand{\rrt}{\ROLE{r}_2}
\newcommand{\rr}{\ROLE{r}}
\newcommand{\Buy}{\ROLE{B}}
\newcommand{\Sell}{\ROLE{S}}
\newcommand{\Agency}{\ROLE{A}}
\newcommand{\DB}{\ROLE{DB}}
\newcommand{\pr}{\alpha}
\newcommand{\prb}{\beta}

\newcommand{\brkin}{|}

\newcommand{\U}{\mode{\GA[\p]}}
\newcommand{\UI}{(\LA[\p])^\Imode}
\newcommand{\UO}{(\LA[\p])^\Omode}
\newcommand{\UIO}{(\GA[\p])^\IOmode}
\newcommand{\UIOG}{\GA[\p]}
\newcommand{\UIA}{\Imode(\assleft A\assright\GA[\p])}
\newcommand{\UA}{\mode{\assleft A\assright\GA[\p]}}
\newcommand{\UOA}{\Omode(\assleft A\assright\GA[\p])}

\newcommand{\VAR}[1]{var(#1)} 
\newcommand{\IVAR}[1]{\mathtt{fv}(#1)}  

\newcommand{\GInter}[2]{#1 \rightarrow #2}
\newcommand{\GForm}[7]{\GInter{#1}{#2}: \ASET{#3 (#4: #5) #6. #7}}
\newcommand{\NGForm}[6]{\GInter{#1}{#2}: \lbrace #3 (#4: #5) #6}
\newcommand{\LFormOut}[6]{#1! \ASET{#2 (#3: #4) #5. #6}}
\newcommand{\LFormIn}[6]{#1? \ASET{#2 (#3: #4) #5. #6}}
\newcommand{\LFormDaggerOne}[6]{#1\dagger_1 \ASET{#2 (#3: #4) #5. #6}}
\newcommand{\LFormDaggerTwo}[6]{#1\dagger_2 \ASET{#2 (#3: #4) #5. #6}}

\newcommand{\LFormOutMarked}[6]{\underline{#1 !} \ASET{#2 (#3: #4) #5. #6}}
\newcommand{\LFormInMarked}[6]{\underline{#1 ?} \ASET{#2 (#3: #4) #5. #6}}

\newcommand{\GInterMarked}[2]{\underline{#1 \rightarrow #2}}
\newcommand{\GFormMarked}[7]{\GInterMarked{#1}{#2}:\ASET{#3 (#4: #5) #6. #7}}

\newcommand{\Rec}[5]{\mu #1 ( #2 ) \ASET{#3} \ENCan{#4} . #5}
\newcommand{\RecDef}[2]{#1 \ENCan{#2}}
\newcommand{\RecDefV}[3]{#1 (#2) \ENCan{#3}}
\newcommand{\GSat}{\mathit{GSat}} 

\newcommand{\typing}[3]{#1\proves #2 \triangleright #3}
\newcommand{\MAssert}[4]{\ENCan{#1, #2, \OT{#3}{#4}}}

\newcommand{\s}{k} 
\newcommand{\sv}{y} 
\newcommand{\sn}{s} 

\newcommand{\aname}{u} 
\newcommand{\av}{x} 
\newcommand{\af}{a} 

\newcommand{\RecT}{\keyword{t}} 

\newcommand{\participant}[1]{\ensuremath{\mathtt{#1}}}
\newcommand{\q}{\ensuremath{\participant{q}}}
\newcommand{\p}{\ensuremath{\participant{p}}}
\newcommand{\player}{\participant{p}}
\newcommand{\ply}{\player} 

\newcommand{\PInter}[2]{#1 [#2]}
\newcommand{\BRANCHcp}[7]{\PInter{#1}{#2,#3}?\{ #4(#5).#6 \}_{#7}}
\newcommand{\BRANCHsig}[6]{\PInter{#1}{#2,#3}? #4(#5).#6}
\newcommand{\Branch}[6]{\{ #1 (#2\! :\!  #3)\{#4\}.#5\}_{#6}}

\newcommand{\SELECTcp}[6]{\PInter{#1}{#2,#3}! #4\ENCan{#5}; {#6}}
\newcommand{\SELECTdef}[5]{\PInter{#1}{#2,#3}! #4\ENCan{#5}}

\newcommand{\AOUTPUT}[4]{\overline{#1}\ENCan{#2[#3]:#4}} 
\newcommand{\AINPUT}[5]{{#1} (\AT{#2[#3]}{#4}). #5} 
\newcommand{\AINPUTT}[4]{{#1} (\AT{#2[#3]}{#4})} 
\newcommand{\ABOUTPUT}[2]{\overline{#1}(#2)}
\newcommand{\ABINPUT}[2]{{#1}(#2)}

\newcommand{\Qin}[2]{#1^\imode\!\!:\!#2}
\newcommand{\Qout}[2]{#1^\omode\!\!:\!#2}
\newcommand{\Qext}[2]{#1\!:\!#2}

\newcommand{\IF}[1]{{\text{if}}\ #1}
\newcommand{\IFTHENELSE}[3]{\text{if}\ #1\ \text{then}\ #2\ \text{else}\ #3}

\newcommand{\new}[3]{\keyword{new}\: \AT{#1}{#2}\ \keyword{in}\ #3}

\newcommand{\newandjoin}{\keyword{new}\ s : \{a_i : \LA_i[\rr_i]\}_{i\in I} \ \keyword{in}\ P}

\newcommand{\SJOIN}[3]{\mathsf{join}\ #1 [#2] ;#3}
\newcommand{\SLEAVE}[3]{\mathsf{leave}\ #1 [#2] ;#3}
\newcommand{\JOIN}[4]{#1 [#2]:#3 . #4}
\newcommand{\JOINP}[3]{\mathsf{join}\ #1\ \mathsf{as}\ #2\ \mathsf{in}\ #3}
\newcommand{\recur}[6]{(\mu #1 (#2\ \ #3).#4)\ENCan{#5, #6}}
\newcommand{\recurDef}[3]{#1 (#2\ \ #3)}
\newcommand{\recurIn}[2]{\ENCan{#1\ \ #2} }

\newcommand{\Nus}[2]{(\nu #1: #2)}
\newcommand{\Nua}[2]{(\nu #1: #2)}

\newcommand{\INACT}{\mathbf{0}}
\newcommand{\inact}{\mathbf{0}}

\newcommand{\proj}{\pmb{\pmb{\upharpoonright}}}
\newcommand{\Proj}[2]{#1 \upharpoonright {#2}}
\newcommand{\pmt}{\kw{pm}}
\newcommand{\PMForm}[3]{ #1 ! #2 (#3)}
\newcommand{\SelfType}[2]{\keyword{st}(#1, #2)}

\newcommand{\LInter}[2]{#1 [#2]}

\newcommand{\AOUT}[4]{\overline{#1}\ENCan{#2[#3]:#4}}
\newcommand{\AIN}[4]{#1\ENCan{#2[#3]:#4}}

\newcommand{\SOUTPUT}[5]{#1[#2,#3]!\OP{#4}{#5}} 
\newcommand{\SINPUT}[5]{#1[#2,#3]?\OP{#4}{#5}} 

\newcommand{\newl}[2]{\keyword{new}\ #1\ :\ #2}

\newcommand{\TRANS}[1]{\xrightarrow{#1}}
\newcommand{\TRANSS}[1]{{\xrightarrow{\raisebox{-.3ex}[0pt][0pt]{\scriptsize $#1$}}}}
\newcommand{\TAUTRANS}[1]{\stackrel{#1}{\Longrightarrow}}
\newcommand{\TAUTRANSh}[1]{\stackrel{\hat{#1}}{\Longrightarrow}}

\newcommand{\CHAIN}[1]{ (#1)^{\keyword{c}}}
\newcommand{\LCHAIN}[2]{ (#1)^{\keyword{c}_{#2}}}
\newcommand{\MCHAIN}[1]{ (#1)^{\keyword{cm}}}
\newcommand{\MLCHAIN}[2]{ (#1)^{\keyword{cm}_{#2}}}
\newcommand{\INCHAIN}[1]{ (#1)^{\keyword{ci}}}
\newcommand{\INLCHAIN}[2]{ (#1)^{\keyword{ci}_{#2}}}

\newcommand{\VECL}{\mathfrak{L}} 

\newcommand{\Prod}[1]{\prod_{#1}}

\newcommand{\GW}{{\mathbb{G}}}
\newcommand{\ACTIVE}[1]{#1} 
\newcommand{\NormalACT}[1]{#1} 
\newcommand{\INACTIVE}[1]{#1} 
\newcommand{\INFER}[2]{\frac{\displaystyle{#1}%
\vspace{2mm}%
}{
\vspace{2mm}%
\displaystyle{#2}
}}

\newenvironment{CASES}
{\left\{%
 \begin{array}{lc}}%
{\end{array}%
 \right.}

\newcommand{\CONFORMS}[2]{#1 \models #2}
\newcommand{\GTRANS}[1]{\TRANS{#1}_\text{g}}
\newcommand{\REL}[1]{\mathcal{#1}}
\newcommand{\RED}{\longrightarrow}

\newcommand{\keyword}[1]{\textsf{\upshape #1}}
\newcommand{\kw}{\keyword} 
\newcommand{\defk}{\keyword{def}}
\newcommand{\ink}{\keyword{in}}

\newcommand{\andk}{\keyword{and}}
\newcommand{\ork}{\keyword{or}}
\newcommand{\NOT}[1]{\;\;\;\not\!\!\!\!\!\!#1}

\newcommand{\truek}{\keyword{tt}}
\newcommand{\falsek}{\keyword{ff}}
\newcommand{\acceptk}{\keyword{accept}}
\newcommand{\requestk}{\keyword{request}}

\newcommand{\ifk}{\keyword{if}}
\newcommand{\thenk}{\keyword{then}}
\newcommand{\elsek}{\keyword{else}}
\renewcommand{\ifthenelse}[3]{\ifk\ #1\ \thenk\ #2\ \elsek\ #3}
\newcommand{\kend}{\keyword{end}}

\newcommand{\ADD}{\mathfrak{+}} 
\newcommand{\AND}{\wedge}
\newcommand{\ANDw}{\,\wedge\,}
\newcommand{\OR}{\vee}
\newcommand{\ORw}{\,\vee\,}
\newcommand{\DELETE}{\mathfrak{-} } 
\newcommand{\Retrv}{\mathfrak{-}}

\newcommand{\fnk}{\keyword{fn}}

\newcommand{\trule}[1]{{\footnotesize{\ensuremath{\lfloor\text{\sc{#1}}\rfloor}}}}
\newcommand{\mrule}[1]{{\footnotesize{\ensuremath{[\text{\sc{#1}}]}}}}

\newcommand{\srule}[1]{{\footnotesize{\ensuremath{(\text{\sc{#1}})}}}}

\newcommand{\monrule}[1]{{\footnotesize{\ensuremath{\lceil{\text{\sc{#1}}}\rceil}}}}

\newcommand{\lprule}[1]{{\footnotesize{\ensuremath{\{\text{\sc{#1}}\}}}}}

\newcommand{\natk}{\kw{nat}}
\newcommand{\boolk}{\kw{bool}}
\newcommand{\intk}{\keyword{int}}
\newcommand{\stringk}{\keyword{string}}

\newcommand{\defink}[2]{\defk\ #1\ \ink\ #2}
\def\eqdef{\;\stackrel{\text{\scriptsize def}}{=}\;}
\def\DEFEQ{\eqdef}

\newcommand{\proves}{\vdash}
\newcommand{\has}{\triangleright}
\newcommand{\IFF}{\Leftrightarrow}
\newcommand{\NULL}{\varepsilon}

\newcommand{\SB}{\sim}
\newcommand{\WB}{\approx}
\newcommand{\LITEQ}{\equiv}              
\newcommand{\Cong}{\equiv} 
\newcommand{\SUBS}[2]{[{#1}/{#2}]}
\newcommand{\MSUBS}[2]{\{{#1}/{#2}\}}
\newcommand{\OFF}[2]{#1 / #2}


\newcommand{\sbj}[1]{\mathsf{sbj}(#1)}
\newcommand{\domain}[1]{\mathsf{dom}(#1)}
\newcommand{\fn}[1]{\mathsf{fn}(#1)}
\newcommand{\bn}[1]{\mathsf{bn}(#1)}

\newcommand{\role}[1]{#1}
\newcommand{\names}[1]{\mathsf{n}(#1)}
\newcommand{\n}[1]{\mathsf{n}(#1)} 
\newcommand{\num}[1]{\mathsf{num}(#1)}

\newcommand{\erase}[1]{\mathsf{erase}(#1)}
\newcommand{\dom}[1]{\mathsf{dom}(#1)}
\newcommand{\complete}[1]{\keyword{COM}(#1)}
\newcommand{\permutableTo}{\curvearrowright}
\newcommand{\permutableToOnce}{\curvearrowright^1}

\newcommand{\LP}{\mathcal{L}}
\newcommand{\RPP}{RPP}
\newcommand{\Location}{\keyword{L}}
\newcommand{\capability}[1]{\mathsf{capability}(#1)}
\newcommand{\CAPABILITY}[1]{\mathsf{cap}(#1)}
\newcommand{\IMAGINARY}{{imaginary}}
\newcommand{\INNETWORK}{\mbox{in the network}} 
\newcommand{\OUTNETWORK}{\mbox{out of the network}} 

\newcommand{\Adaptor}{\mathfrak{A}}
\newcommand{\ini}{\keyword{ini}}

\newcommand{\JOINl}[1]{\mathsf{join}(#1)} 
\newcommand{\JSESS}[1]{#1^\bullet}

\newcommand{\localErr}{\mathsf{localErr}}
\newcommand{\envErr}{\mathsf{envErr}}
\newcommand{\Err}{\mathsf{err}}

\newcommand{\err}{\mathsf{err}}

\newcommand{\EVE}{\mathcal{E}}
\newcommand{\EVEG}{\Theta}
\newcommand{\EVED}{\Delta}
\newcommand{\EveLTS}{\GTRANS}

\newcommand{\pick}[2]{\rightsquigarrow(#1, #2)}

\newcommand{\SBeve}{\sim_{\mathsf{eve}}}
\newcommand{\WBeve}{\approx_{\mathsf{eve}}}

\newcommand{\AOUTNT}[3]{\overline{#1}\ENCan{#2[#3]}}
\newcommand{\NAOUTPUT}[2]{\overline{#1} \ENCan{#2}}
\newcommand{\AOUTPUTNT}[4]{\overline{#1}\ENCan{#2[#3]}; #4} 
\newcommand{\unmark}[1]{\mathsf{unmark}(#1)}

\newcommand{\GAmarked}{\underline{\GA}}  
\newcommand{\LAmarked}{\underline{\LA}}  

\newcommand{\TZ}[1]{#1}
\newcommand{\TZC}[1]{{#1}}

\newcommand{\JOINS}[3]{\kw{join}\ #1\ \kw{as}\ #2:#3}
\newcommand{\MCon}{{C}}

\newcommand{\GF}{\textbf{G}}

\newcommand{\pt}{\kw{pt}}

\newcommand{\namedP}[3]{[#2]_{#3}\mid #1}
\newcommand{\formalM}{\mon{\Gamma; \Delta}@\alpha}
\newcommand{\NP}[2]{[#1]_{#2}}

\newcommand{\namedPmo}[2]{[#1]_{#2}} 

\newcommand{\monitored}[3]{#1[{\mathcal{M}}(#2)\proves #3]}

\newcommand{\field}{\jmath}
\newcommand{\Leffect}[2]{\ENCan{#1\ , #2}}
\newcommand{\Reffect}[2]{\ASET{#1\ , #2}}
\newcommand{\LeffectRole}[3]{\ENCan{#1^{#3}\ , #2^{#3}}}
\newcommand{\ReffectRole}[3]{\ASET{#1^{#3}\ , #2^{#3}}}

\newcommand{\sumtpB}[9]{
#1 \! \rightarrow \! #2: \{ #3 \! (#4)
\! \LEFF{#5} \! \REFF{#7}.#9\}
}

\newcommand{\sumtpBcom}[8]
{
  #1 \! \rightarrow \! #2:
  \{ #3(#4)
  \REFF{{#5}}
  }

\newcommand{\lsumOut}[6]{
#1! \{ #2(#3) \REFF{#4}.#6 \}
}

\newcommand{\lsumIn}[6]{
#1? \{ #2(#3) \REFF{#4}.#6 \}}

\newcommand{\rectp}[5]{(\mu #1(#2)\ASET{#3}.#4)\ENCan{#5}}
\newcommand{\insttp}[2]{#1\ENCan{#2}}

\newcommand{\newname}[1]{\pmb{\nu}\, #1}

\newcommand{\prohibited}[2]{\keyword{prohibitedUnder}(#1, #2)}

\newcommand{\allowed}[2]{\keyword{allowed}(#1, #2)}
\newcommand{\after}[2]{\keyword{after}(#1, #2)}

\newcommand{\joinm}[2]{\mathit{join}\ENCan{#1[#2]}}
\newcommand{\jointype}[3]{\mathit{join}\ENCan{#1[#2]:#3}}
\newcommand{\leavem}[2]{\mathit{leave}\ENCan{#1[#2]}}
\newcommand{\leavetype}[3]{\mathit{leave}\ENCan{#1[#2]:#3}}
\newcommand{\newm}[1]{\mathit{new}\ENCan{#1}}

\newcommand{\newmNPD}[4]{\mathit{new}\ENCan{#1:#2[#3],#4}}

\newcommand{\Message}{M}

\newcommand{\newNP}[4]%
{\mathsf{new}\ #1 \ \mathsf{with} \ #2[#3]\ \mathsf{in} \ #4}

\newcommand{\newNPmo}[3]%
{\mathsf{new}\ #1 \ \mathsf{with} \ #2\ \mathsf{in} \ #3}

\newcommand{\newNPD}[5]%
{\mathsf{new}\ #1 \ \mathsf{with} \ #2[#3], #5\ \mathsf{in} \ #4}

\newcommand{\interfacedN}[2]{#1[#2]}

\newcommand{\AddressInterface}{\CAL{A}}
\newcommand{\AI}{\AddressInterface}
\newcommand{\BI}{\CAL{B}}
\newcommand{\Monitor}{\CAL{M}}

\newcommand{\MonitorG}{\mon{\Gamma}}

\newcommand{\namedM}[2]{{#1}}

\newcommand{\WFN}[2]{\proves \interfacedN{#1}{#2}\,:\,\diamond}
\newcommand{\ADR}[1]{\text{adr}(#1)}

\newcommand{\InputAct}[1]{\Downarrow #1}
\newcommand{\OutputAct}[1]{\Uparrow #1}
\newcommand{\TauAct}{\mathbf{\tau}}
\newcommand{\tauact}{\mathbf{\tau}}

\newcommand{\obj}[1]{\mathsf{obj}(#1)}

\newcommand{\NEWP}[2]{\mathsf{new}\ #1\ \mathsf{in}\ #2}

\newcommand{\newP}[2]{\mathsf{new}\ #1\ \mathsf{in}\ #2}

\newcommand{\destination}[1]{\mathsf{dest}(#1)}
\newcommand{\source}[1]{\mathsf{src}(#1)}
\newcommand{\Incoming}{\mathsf{incoming}}
\newcommand{\Outgoing}{\mathsf{outgoing}}

\newcommand{\INW}[2]{#1\proves #2}

\newcommand{\MAP}[1]{[\![#1]\!]}
\newcommand{\MAPenc}[1]{\ENCan{\!\ENCan{#1}\!}}

\newcommand{\monLeft}{\langle\!\langle}
\newcommand{\monRight}{\rangle \!\rangle}
\newcommand{\mon}[1]{\monLeft #1 \monRight}
\newcommand{\unmon}[1]{\mathsf{unmon}(#1)}


\newcommand{\getBI}[1]{\mathsf{int}(#1)}

\newcommand{\COMMENT}[1]{}
\newcommand{\PREDICATE}[1]{{\cal{#1}}}
\newcommand{\EFFECT}[1]{{\ENCan{#1}}}

\newcommand{\midW}{\,\mid\,}
\newcommand{\midWW}{\ \mid\ }

\newcommand{\coherent}{\asymp}

\newcommand{\GQ}[2]{\ENCan{#1\ ;\ #2}}

\newcommand{\smsg}[5]{#1\ENCan{#2,#3,#4\ENCan{#5}}}

\newcommand{\OUTPUT}[3]{\OL{#1}\ENCan{#2};#3}
\newcommand{\INPUT}[3]{#1(#2).#3}

\newcommand{\hastype}{\triangleright}

\newcommand{\effected}[2]{#1\,\mathtt{after}\,#2}

\newcommand{\DSP}{{\sf DSP}}

\newcommand{\PRG}{{P}}
\newcommand{\PRGQ}{{Q}}
\newcommand{\sN}{M}
\newcommand{\dN}{N}

\newcommand{\PRT}{P}

\newcommand{\PRTone}{P_{rt}}

\newcommand{\RInfo}{r}  

\newcommand{\DECLARE}[1]{(\!(#1)\!)}

\newcommand{\ONW}[2]{{#1}} 

\newcommand{\ENV}{\OL{N}}

\newcommand{\IMPLIES}{\Longrightarrow}

\newcommand{\NWSAT}[3]{#1 \models #2 : #3}
\newcommand{\NWSATa}[3]{#1 \models' #2 : #3}
\newcommand{\TRIPLE}[3]{\ENCan{#1, #2, #3}}

\newcommand{\goodP}[3]{#1\, :_#2\, #3}
\newcommand{\goodNW}[2]{#1\, :\, #2}
\newcommand{\goodNWunder}[3]{#1 \,\triangleright\, \goodNW{#2}{#3}}

\newcommand{\NWgood}[2]{#1 \,\triangleright\, #2}
\newcommand{\NLNWgood}[2]{#1 \,:\, #2}
\newcommand{\Ocompose}[2]{#1 \uplus #2}
\newcommand{\Icompose}[2]{#1 , #2}

\newcommand{\principals}[1]{\mathcal{P}(#1)}

\newcommand{\AEnv}{\mathcal{C}}
\newcommand{\InvExtract}[1]{\mathtt{inv}(#1)}
\newcommand{\AllExtract}[1]{\mathtt{inv}(#1)}
\newcommand{\offsession}[3]{#1[\ROLE{#2}]}
\newcommand{\onsession}[3]{#1[\ROLE{#2}]}
\newcommand{\InvNote}[1]{\tiny\blacksquare_{#1}}
\newcommand{\InvNotes}[1]{\tiny\widetilde{\blacksquare}_{#1}}
\newcommand{\InvA}{\textbf{A}}
\newcommand{\newNPmoPR}[3]%
{\mathsf{new}\ #1 \ \mathsf{with} \ [#2]_{\sigma'}\ \mathsf{in} \ #3}
\newcommand{\SEnv}{\mathcal{S}}
\newcommand{\state}[1]{S^{#1}}
\newcommand{\lstate}[1]{\locked{S}^{#1}}
\newcommand{\ustate}[1]{\unlocked{S}^{#1}}
\newcommand{\statep}[1]{S'^{#1}}
\newcommand{\lstatep}[1]{\locked{S'}^{#1}}
\newcommand{\ustatep}[1]{\unlocked{S'}^{#1}}

\newcommand{\OLtrule}[1]
{{\footnotesize{\ensuremath{\lfloor\OL{\text{\sc{#1}}\rfloor}}}}}

\newcommand{\LINKED}{{dynamic}}
\newcommand{\UNLINKED}{{static}}

\newcommand{\ROLES}[1]{{\mathsf{roles}(#1)}}

\newcommand{\OSPEC}{{\Theta^{\mathsf{ex}}}}
\newcommand{\ISPEC}{{\Theta^{\mathsf{in}}}}

\newcommand{\OSPECi}[1]{{\Theta^{\mathsf{ex}}_#1}}
\newcommand{\ISPECi}[1]{{\Theta^{\mathsf{in}}_#1}}

\newcommand{\NWsat}[3]{#1 \triangleright\, #2 \,:\, #3}

\newcommand{\EC}[1]{{\mathcal E}(#1)}
\newcommand{\EConly}{{\mathcal E}}

\newcommand{\NEC}[1]{{\mathcal N}(#1)}
\newcommand{\NEConly}{{\mathcal N}}

\newcommand{\LockP}{\mathcal{P}}
\newcommand{\LockPOut}{\LockP^\uparrow}
\newcommand{\LockPIn}{\LockP^\downarrow}
\newcommand{\Lock}{\blacktriangledown}
\newcommand{\Unlock}{\blacktriangle}
\newcommand{\Locking}[1]{\Lock #1 \Unlock}
\newcommand{\Evaluate}{\square}
\newcommand{\Read}{\boxdot}
\newcommand{\Write}{\boxtimes}

\newcommand{\GET}{\VEC{x}=get(\VEC{\field});}
\newcommand{\PUT}{put(\VEC{e}, \VEC{\field});}
\newcommand{\PUTUNLOCK}{put(\VEC{e}, \VEC{\field})\Unlock}
\newcommand{\UNLOCK}{\Unlock}

\newcommand{\GETxi}[2]{{#1}=get(#2)}
\newcommand{\PUTxi}[2]{put(#1, #2)}

\newcommand{\Locked}[1]{\underline{#1}}
\newcommand{\LockedInput}[6]
{#1[#2, #3]  ?\Lock \{ #4(#5). #6\}}

\newcommand{\ri}[1]{\mathsf{route}(#1)}
\newcommand{\control}{\mathsf{control}} 
\newcommand{\addinfo}{\varrho} 
\newcommand{\newsLabel}[4]{\keyword{new}(s) \{a_i: \LA_i[\rr_i]\} }
\newcommand{\newaLabel}[3]{\keyword{reg}~a :{#2}[#3]}
\newcommand{\newpLabel}[3]{\nu \alpha :{#2}[#3]} 
\newcommand{\pendingInvitations}[4]{\{a_i\mapsto #1[#2]:#3\}}

\newcommand{\Scribble}{Scribble}
\newcommand{\PARA}[1]{\paragraph*{{\bf #1.}}}

\newcommand{\Ser}{\texttt{S}}
\newcommand{\Cli}{\texttt{C}}
\newcommand{\Agent}{\texttt{A}}
\newcommand{\FON}[1]{{\mathtt{fobj}}(#1)} 

\newcommand{\nEC}[1]{{#1}}

\newcommand{\dual}{\mathit{dual}}





\newcommand{\M}{\mathsf{M}}
\newcommand{\Mp}{\mathsf{M}'}

\newcommand{\monitor}{\M}

\newcommand{\A}{\mathcal{M}^{\circ}}
\newcommand{\Ap}{\mathcal{M}_2^{\circ}}


\newcommand{\Moff}[1]{\monitor^{\circ}[#1]}
\newcommand{\Moffone}[1]{\monitor_1^{\circ}[#1]}
\newcommand{\Mofftwo}[1]{\monitor_2^{\circ}[#1]}
\newcommand{\Moffp}[1]{\monitor_2^{\circ}[#1]}
\newcommand{\Mon}[1]{\monitor[#1]}
\newcommand{\Monp}[1]{\monitor'[#1]}
\newcommand{\Monone}[1]{\monitor_1[#1]}
\newcommand{\Montwo}[1]{\monitor_2[#1]}

\newcommand{\ptp}[1]{{\participant{#1}}}

\newcommand{\pf}{\__{T}}

\newcommand{\Recursion}[4]{\mu \keyword{t}.#3}
\newcommand{\RecursionT}[3]{\mu \keyword{t}\ENCan{#1}(#2).#3}
\newcommand{\RecursionP}[3]{\RecursionPAbs{#1}{#2}}
\newcommand{\RecursionPAbs}[2]{\mu X.#2}
\newcommand{\CallT}[1]{\keyword{t}}
\newcommand{\Call}[1]{\keyword{t}}
\newcommand{\CallP}[1]{X}
\newcommand{\typeconst}[1]{\mathsf{nat}}
\newcommand{\nat}{\typeconst{nat}}
\newcommand{\intTp}{\textsf{Int}}
\newcommand{\bool}{\textsf{Bool}}

\newcommand{\NTRANS}[1]{\stackrel{#1}{\not\longrightarrow}}
\newcommand{\const}[1]{\mathtt{#1}}
\newcommand{\ENTAILS}{\supset}
\newcommand{\MPSA}{{\it MPSA}}
\newcommand{\MPST}{{\it MPST}}

\newcommand{\MPSTs}{{\it MPST}s}
\newcommand{\MPSAs}{{\it MPSA}s}

\newcommand{\eveproves}{\proves_{\text{g}}}

\newif\ifny\nytrue
\nytrue
\newcommand{\NY}[1]
	{\ifny{\color{violet}{#1}}\else{#1}\fi}
\newcommand{\KH}[1]
	{\ifny{\color{red}{#1}}\else{#1}\fi}
\newcommand{\TZU}[1]
	{\ifny{\color{blue}{#1}}\else{#1}\fi}
\definecolor{dkgreen}{rgb}{0,0.5,0}
\newcommand{\MD}[1]
	{\ifny{\color{dkgreen}{#1}}\else{#1}\fi}
\newcommand{\RH}[1]
	{\ifny{\color{orange}{#1}}\else{#1}\fi}

\newcommand{\GC}{\mathcal{G}}
\newcommand{\LC}{\mathcal{T}}

\newcommand{\grmor}{\ \mathrel{\text{\large$\mid$}}\ }

\newcommand{\MPSAenv}{\envT{\Gamma}{\Delta}{D}}
\newcommand{\SPECenv}{\Theta}

\newcommand{\envTriple}[3]{#1;\, #2 #3}
\newcommand{\envT}[3]{\envTriple{#1}{#2}{#3}}
\newcommand{\envTT}[3]{\ENCan{\envTriple{#1}{#2}{#3}}}

\newcommand{\envQuintuple}[5]{#1;\, #2;\, #3;\,#4;\, #5}
\newcommand{\envQ}[5]{\envQuintuple{#1}{#2}{#3}{#4}{#5}}
\newcommand{\envQQ}[5]{\ENCan{\envTriple{#1}{#2}{#3}{#4}{#5}}}


\newcommand{\LNW}[4]{(\nu #1)(#2;\ \GQ{#3}{#4})}
\newcommand{\LNWn}[3]{(\nu \VEC{n})(#1;\ \GQ{#2}{#3})}

\newcommand{\ULNW}{\kw{N}}

\newcommand{\ULspec}{\kw{\Theta}}

\newcommand{\monitoredP}[3]{#1[#2]_{#3}}

\newcommand{\nubf}{{\pmb \nu}}

\newcommand{\MRED}{\rightarrow\!\!\!\!\!\rightarrow}

\newcommand{\AssEnv}{\ENCan{\Gamma;\Delta}}

\newcommand{\static}[1]{#1}
\newcommand{\dynamic}[1]{#1}

\newcommand{\sta}[1]{\static{#1}}
\newcommand{\dyn}[1]{\dynamic{#1}}

\newcommand{\codom}[1]{\mathsf{codom}(#1)}
\newcommand{\SHS}{Checkability}
\newcommand{\sHS}{checkability}
\newcommand{\STS}{TS}
\newcommand{\WTS}{WTS}

\newcommand{\TSWI}[3]{\STS(#1, #2, #3)}

\newcommand{\TSG}[2]{\WTS(#1)_{#2}}
\newcommand{\TSGI}[3]{\WTS(#1)_{#2}^{#3}}
\newcommand{\HSEnv}{{E}}
\newcommand{\intVar}[1]{\#(#1)}
\newcommand{\stVar}[1]{sv(#1)}
\newcommand{\stateVar}[1]{\mathtt{split_{stateVar}}(#1)}
\newcommand{\inv}[1]{\mathtt{inv}(#1)}
\newcommand{\svar}[1]{\mathtt{svar}(#1)}
\newcommand{\MYSTATEVAR}[2]{\kw{#2}}
\newcommand{\dec}{\mathbf{\mathtt{d}}}
\newcommand{\romain}[1]{$\mathbf{RD}$: #1 $\bigstar$}
\newcommand{\many}[1]{\overtilde{#1}}
\newcommand{\II}[1]{\mathtt{I}(#1)}
\newcommand{\OO}[1]{\mathtt{O}(#1)}

\newcommand{\SERVERROLE}{\texttt{S}}
\newcommand{\CLIENTROLE}{\texttt{C}}
\newcommand{\rumi}[2]{$\mathbf{RN}$:#1 \textcolor{gray}{#2} $\bigstar$}
\definecolor{dkblue}{rgb}{0,0.1,0.5}
\definecolor{dkgreen}{rgb}{0,0.4,0}
\definecolor{dkred}{rgb}{0.4,0,0}

\lstnewenvironment{PYTHONLISTING}%
{
\lstset{
  language=python,
  showstringspaces=false,
  formfeed=\newpage,
  tabsize=2,
  commentstyle=\color{dkgreen}\itshape,
  basicstyle=\ttfamily,
  morekeywords={models, lambda, forms, def, class}
  keywordstyle=\color{dkblue},
  emph={access,and,as,break,class,continue,def,del,elif,else,%
	except,exec,finally,for,from,global,if,import,in,is,%
	lambda,not,or,pass,print,raise,return,try,while,assert,with},
  emphstyle=\color{dkblue}\bfseries,
  basicstyle=\CODESTYLE\footnotesize,
  keywordstyle=\CODESTYLE\footnotesize, 
  escapeinside={*@}{@*}
}
}
{
}

\newcommand{\rtsyntax}[1]{\colorbox{lightgray}{\ensuremath{#1}}}
\newcommand{\upd}{\mathit{update}}
\newcommand{\eq}{\ensuremath{\diameter}}
\newcommand{\kf}[1]{\textup{\textsf{#1}}\xspace}
\newcommand{\constf}[1]{\textup{\textsf{#1}}}
\newcommand{\srsimple}[3]{\ensuremath{\bar{#1}[#2](#3)}}
\newcommand{\sr}[4]{\ensuremath{\srsimple{#1}{#2}{#3}.#4}}
\newcommand{\uu}{\ensuremath{u}}
\newcommand{\Ia}{\ensuremath{a}}
\newcommand{\Ic}{\ensuremath{c}}
\newcommand{\Ias}{\ensuremath{\alpha}}
\newcommand{\Ib}{\ensuremath{b}}
\newcommand{\y}{\ensuremath{y}}
\newcommand{\PP}{\ensuremath{P}}
\newcommand{\Q}{\ensuremath{Q}}
\newcommand{\R}{\ensuremath{R}}
\newcommand{\h}{\ensuremath{h}}
\newcommand{\queue}{\ensuremath{\h}}
\newcommand{\DD}{\ensuremath{D}}
\newcommand{\sub}[2]{\ensuremath{\{#1/#2\}}}
\newcommand{\sasimple}[3]{\ensuremath{#1[#2](#3)}}
\newcommand{\sa}[4]{\ensuremath{\sasimple{#1}{#2}{#3}.#4}}
\newcommand{\redsym}{\ensuremath{\longrightarrow}}
\newcommand{\nm}[1]{{\texttt{\scriptsize #1.}}}
\newcommand{\ty}{\textbf{t}}
\newcommand{\End}{\kf{end}}
\newcommand{\pro}[2]{\ensuremath{#1\upharpoonright#2}}
\newcommand{\G}{\ensuremath{G}}
\newcommand{\Gv}[4]{\ensuremath{#1\to\pset:\langle#3\rangle.#4}}
\newcommand{\PBig}{\ensuremath{P}}
\newcommand{\red}[2]{\ensuremath{#1\redsym#2}}
\newcommand{\X}{\ensuremath{X}}
\newcommand{\x}{\ensuremath{x}}
\newcommand{\pset}{\ensuremath{\Pi}}
\newcommand{\sI}[1]{\ensuremath{\s_{#1}}}
\newcommand{\pI}[1]{\ensuremath{\PBig_{#1}}}
\newcommand{\kk} {\ensuremath{\kappa}}
\newcommand{\si}[2]{\ensuremath{#1[#2]}}
\newcommand{\ki}[2]{\ensuremath{#1(#2)}}
\newcommand{\valheap}[3]{\ensuremath{( #3,\pset,#1 )}}
\newcommand{\va}{\ensuremath{v}}
\newcommand{\sd}[4]{\ensuremath{#1!\langle\! \langle#3,#2\rangle \!\rangle;#4}}
\newcommand{\delheap}[3]{\ensuremath{(#3,{#2},#1)}}
\newcommand{\labheap}[3]{\ensuremath{( #3,\pset,#1 )}}
\newcommand{\ptilde}[1]{{\ensuremath{#1}}}
\newcommand{\at}[1]{\ensuremath{\ptilde{#1}}}
\newcommand{\proccalldots}[3]{\ensuremath{#1\langle\ptilde{#2},\ptilde{#3}\rangle}}
\newcommand{\Xsignature}{\ensuremath{\X(\at{x}, \at{y})}}
\newcommand{\Ddef}{\ensuremath{\Xsignature=\PP}}
\newcommand{\defX}{\ensuremath{\kf{def} \ \Ddef\ \kf{in}\ }}
\newcommand{\anglep}[2]{\ensuremath{\langle #1, #2\rangle}}
\newcommand{\set}[1]{\ensuremath{\{#1\}}}
\newcommand{\T}{\ensuremath{T}}
\newcommand{\UT}{\ensuremath{U}}
\newcommand{\oT}[2]{\ensuremath{\;!\langle #2,#1\rangle}}
\newcommand{\seltype}{\ensuremath{\oplus \langle \pset,\{l_i:\T_i\}_{i\in
I} \rangle }}
\newcommand{\Par}{\ensuremath{\ |\ }}
\newcommand{\pc}{\Par}
\newcommand{\qtail}[1]{\ensuremath{\qcomp{\queue}{#1}}}
\newcommand{\qhead}[1]{\ensuremath{\qcomp{#1}{\queue}}}
\newcommand{\mqueue}[2]{\ensuremath{#1 : #2}}
\newcommand{\qcomp}[2]{\ensuremath{#1 \cdot #2}}
\newcommand{\Th}{\ensuremath{\Theta}}
\newcommand{\dere}[2]{\ensuremath{\Th\,\vdash #1\; \blacktriangleright\; #2 \,;\,\M\,;\,\B}}

\newcommand{\der}[3]{\ensuremath{#1\vdash#2\triangleright#3}}
\newcommand{\Ga}{\ensuremath{\Gamma}}
\newcommand{\D}{\ensuremath{\Delta}}
\newcommand{\sid}[1]{\ensuremath{\textup{pn}(#1)}}
\newcommand{\de}[3]{\ensuremath{#1\vdash#2:#3}}
\newcommand{\ST}{\ensuremath{S}}
\newcommand{\pn}{\p}
\newcommand{\an}[1]{\ensuremath{\langle #1\rangle}}
\newcommand{\ins}{\ensuremath{:}}
\newcommand{\B}{\ensuremath{\mathcal{B}}}
\newcommand{\dereb}[3]{\ensuremath{\Th\,\vdash #1 \;\blacktriangleright\; #2 \,;\,\M\,;\,#3}}
\newcommand{\Or}{\ensuremath{\mathcal{R}}}
\newcommand{\nk}[1]{\ensuremath{\ell(#1)}}
\newcommand{\out}[4]{\ensuremath{#1!\langle \pset,#2\rangle;#4}}
\newcommand{\outp}[3]{\ensuremath{#1!\langle \pset,#2\rangle}}
\newcommand{\outs}[4]{\ensuremath{#1!\langle #3,#2\rangle;#4}}
\newcommand{\e}{\ensuremath{e}}
\newcommand{\adde}[2]{\ensuremath{#1\bar{\cup}\set{#2}}}
\newcommand{\cc}{\ensuremath{c}}
\newcommand{\prule}[1]{\{\text{\textup{\sc{#1}}}\}}
\newcommand{\pre}[2]{\ensuremath{\mathsf{pre}({#1},{#2})}}
\newcommand{\inp}[4]{\ensuremath{#1?( #3,#2);#4}}
\newcommand{\iT}[2]{\ensuremath{?( #2,#1 )}}
\newcommand{\rd}[4]{\ensuremath{#1?(\!(#3,#2)\!);#4}}

\newcommand{\GAs}{G} 

\lstset{ %
language=Java ,                
basicstyle=\footnotesize,       
backgroundcolor=\color{white},  
showspaces=false,               
showstringspaces=false,         
showtabs=false,                 
frame=single,                   
tabsize=2,                      
captionpos=b,                   
breaklines=true,                
breakatwhitespace=false,        
title=\lstname,                 
escapeinside={\%*}{*)},         
morekeywords={*,...}            
}

\title{Multiparty Session Actors}
\author{Rumyana Neykova \quad\quad Nobuko Yoshida
\institute{Imperial College London}
}

\def\titlerunning{Multiparty Session Actors}
\def\authorrunning{R. Neykova \& N. Yoshida}

\maketitle
\begin{abstract}
Actor coordination armoured with a suitable protocol description language has been a pressing problem in the actors community. We study the applicability of multiparty session type (MPST) protocols for verification of actor programs. We incorporate sessions to actors by introducing minimum additions to the model such as the notion of actor roles and protocol mailbox. The framework uses Scribble, which is a protocol description language based on multiparty session types. Our programming model supports actor-like syntax and runtime verification mechanism guaranteeing type-safety and progress of the communicating entities. An actor can implement multiple roles in a similar way as an object can implement multiple interfaces. Multiple roles  allow for inter-concurrency in a single actor still preserving its progress property.  We demonstrate our framework by designing and implementing a session actor library in Python and its runtime verification mechanism.
\end{abstract}

\label{sec:intro}

\section{Introduction}

The actor model is (re)gaining attention in the research community and in the mainstream programming languages as a promising concurrency paradigm. 
Unfortunately, in spite of the importance of message passing mechanisms in the actor paradigm, the programming model itself does not ensure correct sequencing of interactions between different computational processes. 
This is a serious problem when designing safe concurrent and distributed systems in languages with actors. 



To overcome this problem, we need to solve several shortcomings existing in 
the actor programming models. First, although actors often have
multiple states and complex policies for changing states, no
general-purpose specification language is currently in use for
describing actor protocols. Second, a clear guidance on actor
discovery and coordination of distributed actors is missing. As a
study published in {\cite{TasharofiDJ13} reveals, this leads to adhoc
implementations and mixing the model with other paradigms which
weaken its benefits. Third, no verification mechanism (neither
static nor dynamic) is proposed to ensure correct sequencing of
actor interactions. Most actor implementations provide static typing
within a single actor, but the communication between actors -- the
complex communication patterns that are most likely to deadlock --
are not checked.

We tackle the aforementioned challenges by studying applicability of
multyparty session types (MPST) verification and its practical
incarnation, the protocol description language Scribble, to actor
systems. Recent works from \cite{RVTool,RVPaper} prove suitability of
MPST for dynamic verification of real world complex protocols
\cite{OOI}. The verification mechanism is applied to large
cyberinfrastructure, but checks are restricted only to MPST
primitives. 
In this paper, we take MPST verification one step further and apply it
to an actor model by extending it with creation and management of
communication contexts (protocols).

Our programming model is grounded on three design ideas: (1) use
Scribble protocols and their relation to finite state machines for
specification and runtime verification of actor interactions; (2) augment actor
messages and their mailboxes dynamically with protocol (role)
information; and (3) propose an algorithm based on virtual
routers (protocol mailboxes) for the dynamic discovery of actor
mailboxes within a protocol. We implement a session actor
library in Python to demonstrate the applicability of the approach.
To the best of our knowledge, 
this is the first design and
implementation of session types and their dynamic verification
toolchain in an actor
library. 

\section{Multiparty Session Actor Programming}
\subsection{Overview of Multiparty Session Actor Framework}
\begin{figure}
\centering
\includegraphics[scale=0.5]{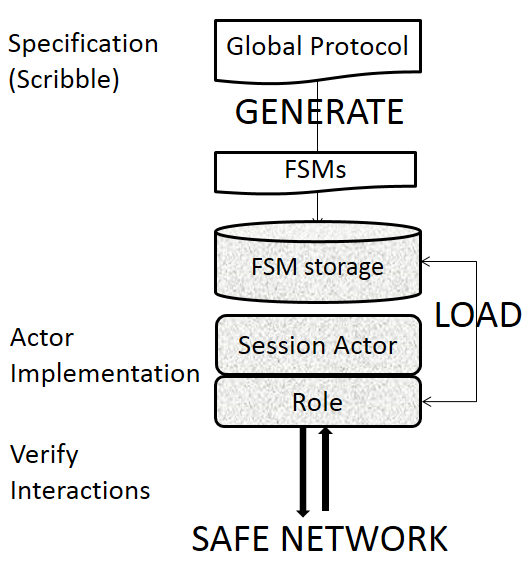}
\vspace{-10pt}
\caption{Development methodology\label{fig:framework}}
\vspace{-10pt}
\end{figure} 

The standard development methodology for MPST verification is
illustrated in Fig.~\ref{fig:framework}, and the contributions of this
work are darked. Distributed protocols are specified in
Scribble, which collectively defines the admissible communication
behaviours. Scribble tool can algorithmically project any global
conversation to the specifications of its endpoints, generating finite
state machines (FSMs). This is different from previous works \cite{RVTool,RVPaper}, where Scribble compiler produces local Scribble protocols and FSMs are
generated on the fly at runtime. This difference is important 
for the applicability of 
our framework to general actor programming. 

Distributed components are realised as session actors associated with
one or more of the generated FSMs. The association is done through
annotating the actor type (class) and receive messages (methods) with
protocol information. Annotating the actor type with protocol
information results in registering the type for a particular role.
When a session is started, a join message is sent (in a round-robin
style) to all registered actors. When a join message is received, the
generated FSM is loaded into the actor role and all subsequent messages on
that protocol (role) are tracked and checked. Message receive is
delegated to the appropriate FSM via pattern matching on
the protocol id, contained in the message. If all actors messages
comply to their assigned FSMs, the whole communication is guaranteed
to be safe. If participants do not comply, violations
(such as deadlocks, communication mismatch or breakage of a protocol)
are detected and delegated to a Policy actor.

\subsection{Warehouse Management Protocol in Session Actors}
To illustrate and motivate central design decisions of our model, we
present the buyer-seller protocol from \cite{HYC08} and extend it to a
full warehouse management scenario. A warehouse consists of multiple
customers communicating to a warehouse provider. It can be involved in
purchase protocol (with customers), but can also be involved in a
loaded protocol with dealers to update its storage. 

\textbf{Scribble Protocol.} \ 
Fig.~\ref{scribble_global} shows the
interactions between the entities in the system written as a Scribble protocol.
There are purchase and storeload protocols, 
involving three (a Buyer~(B), a Seller~(S)
and an Authenticator~(A)) and two (a Store~(S), a Dealer~(D)) parties,
respectively.
  At the start of a purchase session, B
sends login details to S, which delegates the request to an
Authentication server. After the authentication is completed, B sends
a request quote for a product to S and S replies with the product
price. Then B has a choice to ask for another product, to proceed
with buying the product, or to quit. By buying a product the warehouse
decreases the product amount it has in the store. When a product is out of
stock, the warehouse sends a product request to a dealer to load the
store with $n$ numbers of that product (following the storeload protocol). The reader can refer 
to \cite{scribble} for the full specification of 
Scribble syntax. 


{\bf Challenges. }\ 
There are several challenging points to implement the above scenario. First, a
warehouse implementation should be involved in both protocols,
therefore it can play more than one role. Second, initially the user
does not know the exact warehouse it is buying from, therefore the
customer learns dynamically the address of the warehouse. Third, there
can be a specific restriction on the protocol that cannot be expressed
as system constants (such as specific timeout depending on the
customer). The next section explains implementations of Session Actors in more details. 

\begin{figure}[t]	
\begin{minipage}[t]{0.52\linewidth}
\lstset{numbers=left}
\begin{PYTHONLISTING}
global protocol Purchase
	(role B, role S, role A)
{
	login(string:user) from B to S;
	login(string:user) from S to A;
	authenticate(string:token) from A to B, S;
	choice at B
		{request(string:product) from B to S;
		(int:quote) from S to B;}
	or
		{buy(string:product) from B to S
		 delivery(string) from S to B; }
	or 
		{quit() from B to S; }}
global protocol StoreLoad
	(role D, role S)
{	
	rec Rec{
	choice at S
		{request(string:product, int:n) from S to D;)
		 put(string:product, int:n) from D to S;
		 continue Rec;}
	or  
		{quit() from S to D;
		 acc() from D to S;}}}	
\end{PYTHONLISTING}
\vspace{-7pt}
\caption{Global protocols in Scribble}
\label{scribble_global}
\vspace{-10pt}
\end{minipage}
\begin{minipage}[t]{0.49\linewidth}
\begin{PYTHONLISTING}
@protocol(c, Purchase, seller, buyer, auth)*@\label{l:p1}@*
@protocol(c1, StoreLoad, seller, dealer)*@\label{l:p2}@*
class Warehouse(SessionActor):

	@role(c, buyer)*@\label{l:rb}@*
	def login(self, user): *@\label{l:l}@*
		c.auth.send.login(user)
	
	@role(c, buyer)
	def buy(self, product):*@\label{l:bs}@*
		self.purchaseDB[product]-=1;
		c.seller.send.delivery(product.details)*@\label{l:d}@*
		self.become(update, product)*@\label{l:be}@*
	
	@role(c, buyer)
	def quit(self):
		c.send.buyer.acc()	
	
	@role(c1, self)		
	def update(self, product):*@\label{l:us}@*
		c1.dealer.send.request(product, n)*@\label{l:ue}@*		
		
	@role(c1, dealer)
	def put(self, c, product): *@\label{l:p}@*
		self.purchaseDB[product]+=1:	
		
\end{PYTHONLISTING}
\vspace{-7pt}
\caption{Session Actor for warehouse}
\label{master_role_python}
\vspace{-10pt}
\end{minipage}
\end{figure}
\textbf{Session Actor.} \ 
Fig.~\ref{master_role_python} presents an
implementation of a warehouse service as a single session
actor that keeps the inventory as a state ({\it self.purchaseDB}). Lines~\ref{l:p1}-~\ref{l:p2} annotate the
session actor class with two protocol decorators -- {\it c} and {\it c1} (for
seller and store roles respectively). {\it c} and {\it c1} are accessible
within the warehouse actor and are holders for mailboxes of the
other actors, involved in the two protocols. All message handlers are
annotated with a role and for convenience are implemented as methods.
For example, the login method (Line~\ref{l:l}) is invoked when a {\it login}
message (Line 4, Fig.~\ref{scribble_global}) is sent. The role
annotation for {\it c} (Line~\ref{l:rb}) specifies the sender to be {\it buyer}. The
handler body continues following Line 5, Fig.~\ref{scribble_global} -
sending a {\it login} message via the {\it send} primitive to the session
actor, registered as a role {\it auth} in the protocol of {\it c}. Value {\it c.auth}
is initialised with the {\it auth} actor mailbox as a result of the
actor discovery mechanism (explained in the next section). The
handling of {\it authenticate} (Line 6, Fig.~\ref{scribble_global}) and
{\it request} (Line 8, Fig.~\ref{scribble_global}) messages is similar, so
we omit it and focus on the {\it buy} handler (Line~\ref{l:bs}-~\ref{l:be}), where after sending
the delivery details (Line~\ref{l:d}), the warehouse actor sends a message to
itself (Line~\ref{l:be}) using the primitive {\it become} with value {\it update}.
Value {\it update} is annotated with another role {\it c1}, but has as a sender
{\it self}. This is the mechanism used for switching between roles within
an actor. Update method ({Line~\ref{l:us}-~\ref{l:ue}, Fig.~\ref{master_role_python}) implements the
request branch (Line 20-22, Fig.~\ref{scribble_global}) of the
{\it StoreLoad} protocol - sending a request to the {\it dealer} and handling
the reply via method {\it put}.
The correct order of messages is verified by the FSM attached to {\it c} and {\it c1}. As a result, errors such as calling {\it put} before {\it update} or executing two consecutive updates, will be detected as invalid. 
\section{Implementation of Multiparty Session Actors}
\textbf{AMQP in a Nutshell}
\textbf{Distributed actor library.}\ 
We have implemented the multiparty session actors on top of Celery \cite{celery} (distributed message queue in Python) with support for distributed actors. Celery uses advanced message queue protocol (AMQP 0-9-1 \cite{AMQP}) as a transport. The reason for choosing AMQP network as base for our framework is that AMQP middleware shares a similar abstraction with the actor programming model, which makes the implementation of distributed actors more natural. AMQP model can be summarised as follow: messages are published by producers to entities, called exchanges (or mailboxes). Exchanges then distribute message copies to queues using rules called bindings. Then AMQP brokers (virtual routers) deliver messages to consumers subscribed to queues. Distributed actors are naturally represented in this context using the abstractions of exchanges. Each actor type is represented in the network as an exchange and is realised as a consumer subscribed to a queue based on a pattern matching on the actor id. Message handlers are implemented as methods on the actor class.

\begin{figure}[t]
\centering
\includegraphics[scale=0.35]{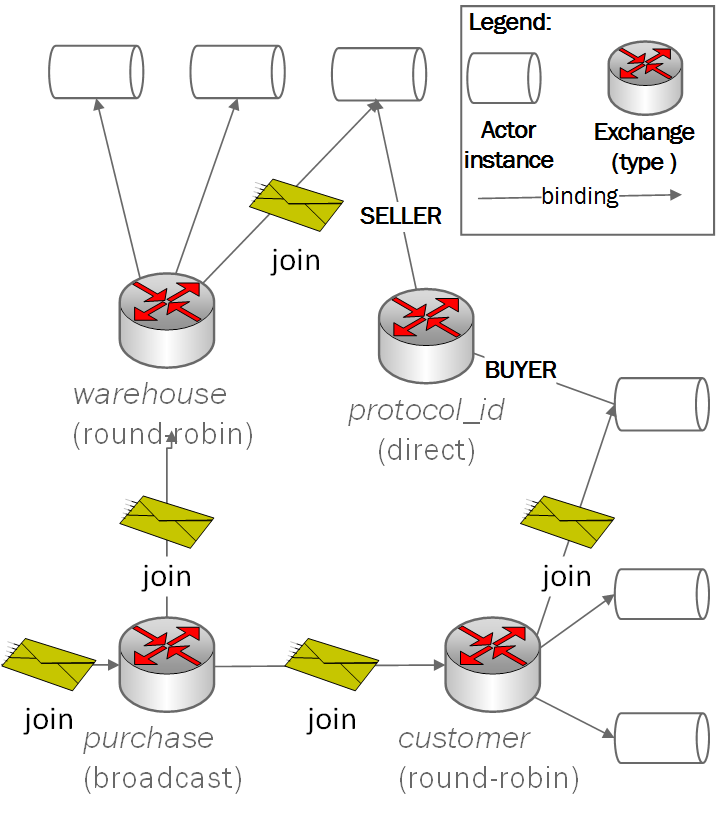}
\vspace{-10pt}
\caption{Organising Actors into protocols\label{fig:actor_into_protocol}}
\label{network}
\vspace{-10pt}
\end{figure}

Our distributed actor discovery mechanism draws on the AMPQ abstractions of exchanges, queues and binding, and our extensions to the actor programming model are built using Python advanced abstraction capabilities: two main capabilities are coroutines (for realising the actors inter-concurrency) and decorators (for annotating actor types and methods). 

\textbf{Actor roles.}\ 
A key idea is each role to run in a virtual thread of an actor (using Python coroutines/green threads). We annotate methods, implementing part of a protocol, with a role decorator. Roles are scheduled cooperatively. This means that at most one role can be active in a session actor at a time. A role is activated when a message is received and ready to be processed. Switching between roles is done via the {\it become} primitive (as demonstrated in Fig.~\ref{master_role_python}), which is realised as sending a message to the internal queue of the actor.  

\textbf{Actors discovery.} \ 
Fig.~\ref{network} presents the network setting (in terms of AMQP objects) for realising the actor discovery for {\it buyer} and {\it seller} of the protocol {\it Purchase}.
For simplicity, we create the actor exchanges on starting of the actor system -- round-robin exchange per actor type ({\it warehouse} and {\it customer} in Fig.~\ref{network}) and broadcast exchange per protocol type ({\it purchase} in Fig.~\ref{network}). All spawned actors alternate to receive messages addressed to their type exchange. Session actors are registered for roles via the protocol decorator and as a result their type exchange is bound to the protocol exchange (Line~\ref{l:p1} in Fig.~\ref{master_role_python} binds {\it warehouse} to {\it purchase} in Fig.~\ref{network}).

We explain the workflow for actor discovery. When a protocol is started, a fresh protocol id and an exchange with that id are created.  
The type of the exchange is {\it direct}\footnote{A direct type means that messages with routing key R are delivered to actors linked to the exchange with binding R.}
 ({\it protocol id} in Fig.~\ref{network}). Then {\it join} message is sent to the protocol exchange and delivered to one actor per registered role ({\it join} is  broadcasted to {\it warehouse} and {\it customer} in Fig.~\ref{network}). On {\it join}, an actor binds itself to the {\it protocol id} exchange with subscription key equal to its role (bindings {\it seller} and {\it buyer} in Fig.~\ref{network}). When an actor sends a message to another actor within the same session (for example {\it c.buyer.send} in Fig.~\ref{master_role_python}), the message is sent to the protocol id exchange (stored in {\it c}) and from there delivered to the {\it buyer} actor. 

\textbf{Order preservation through FSM checking.} \ Whenever a message is received the actor internal loop dispatches the message to the role the message is annotated with. The FSM, generated from the Scribble compiler (as shown in Fig.~\ref{fig:framework}), is loaded when an actor joins a session and messages are passed to the FSMs for checking before being dispatched to their handler. 
The FSM perform checks message labels (already part of the actor payload) and sender and receiver roles (part of the message binding key due to our extension). An outline of the monitor implementation can be also found in \cite{RVPaper}.
The monitor mechanism is an incarnation of the session monitor in \cite{RVPaper} within actors.  




\section{Related and Future Work}
\label{sec:related}	
There are several theoretical works that have studied the
behavioural types for verifying actors \cite{MostrousV11,
  Crafa}. The work \cite{Crafa} proposes 
a behavioural typing system for an actor calculus 
where a type describes a sequence
of inputs and outputs performed by the actor body, while
\cite{MostrousV11} studies session types for a core of Erlang.  
On the practical side, 
the work \cite{ARC06} proposes a framework of three layers for actor 
programming - actors,
roles and coordinators, which resembles roles and
protocol mailboxes in our setting. Their specifications focus on QoS
requirements, while our aim is to describe and ensure 
correct patterns of interactions (message passing). 
Scala actor library \cite{AKKA} (studied in
\cite{TasharofiDJ13}) supports FSM verification mechanism (through inheritance)
and typed channels. Their channel typing is simple so that 
it cannot capture structures of communications such as sequencing,
branching or recursions. These structures ensured by session types are the key
element for guaranteeing deadlock freedom between multiple actors. 
In addition, in \cite{AKKA}, channels and FSMs are
unrelated and cannot be intermixed; 
on the other hand, in our approach, we rely on
external specifications based on the choreography (MPST) and the FSMs usage
is internalised (i.e.~FSMs are automatically generated from a global type),
therefore it does not affect program structures. To our best
knowledge, no other work is linking FSMs, actors and choreographies in
a single framework.

As a future work, we plan to develop (1) extensions to other actor
libraries to test the generality of our framework and (2) extensive evaluations
of the performance overhead of session actors. The initial micro
benchmark shows the overhead of the three main additions to our framework (FSM checking,
actor type and method annotation) is negligible 
per message, see
\cite{OnlineAppendix} (for example, 
the overhead of FSM checking is less than 2\%).  
As actor languages and frameworks are getting more attractive, 
we believe that our framework would become useful for coordinating large-scale, distributed actor programs.

\bibliographystyle{eptcs}{
\footnotesize
\bibliography{session}
}

\end{document}